\documentclass{aa501}
\usepackage{psfig}

\newcommand{\Lsun} {L$_\odot$}
\newcommand{\Msun} {M$_\odot$}

\begin{document}


\title{The Infrared Continuum Radiation of NGC1808\thanks{Based on
observations with ISO (Kessler et al. 1996), an ESA project with
instruments funded by ESA Member States (especially the PI countries:
France, Germany, the Netherlands and the United Kingdom) with the
participation of ISAS and NASA.}}  

\subtitle{A PAH and Polarisation Study}

\author {R.~Siebenmorgen\inst{1}
	\and E.~Kr\"ugel\inst{2}
	\and R.J.~Laureijs\inst{3}}

\institute{
 	European Southern Observatory, Karl-Schwarzschildstr. 2, 
	D-85748 Garching b.M\"unchen, Germany 
\and 
 	Max-Planck-Institut for Radioastronomy, Auf dem H\"ugel 69,
	Postfach 2024, D-53010 Bonn 
\and
	ISO Data Centre, Villafranca del Castillo,  
	P.O. Box 50727, E-28080 Madrid }

\offprints{rsiebenm@eso.org}

\date{Received June 01, 2001 / Accepted Month XX, 2001}

\abstract{The paper is devoted to the understanding of the infrared
emission of nuclear regions in galaxies.  {\it a)} ISO data of
NGC~1808 are presented: spectro--photometry from 5.1 to 16.4 $\mu$m; \
a $25''\times 25''$ map at 6$\mu$m and 170$\mu$m photometry. {\it b)}
The data are complemented by a polarization measurement at 170$\mu$m
(2.5${\pm}$0.4\% at position angle 94\degr${\pm}$5\degr) and a map at
6$\mu$m.  In the map, the degree of polarisation goes up to 20\% in
the outer regions.  We argue that the polarisation is produced by
emission of big grains and exclude very small grains and PAHs or
scattering and extinction.  {\it c)} The mid infrared spectrum shows,
beside the main emission bands, a so far unknown plateau of PAH
features in the $\geq 13\,\mu$m region.  {\it d)} The total spectrum
can be fit under the assumption of optically thin emission.  However,
such a model fails to reproduce the 25$\mu$m point and implies that
the mid infrared is due to very small grains and PAHs.  These
particles would then also have to be responsible for the 6$\mu$m
polarisation, which is unlikely.  {\it e)} To avoid these
difficulties, we successfully turn to a radiative transfer model whose
major feature is the existence of {\it hot spots} produced by the dust
clouds around OB stars.  We demonstrate the decisive influence on the
mid infrared spectrum of both the PAHs and the hot spots.
	}

\maketitle

\section{Introduction}
The nucleus of a galaxy is usually deeply embedded in dust and can
therefore only be probed by infrared (IR) observations.  In this
paper, we present an infrared polarisation and a spectro--photometric
study of the starburst galaxy NGC1808.  The new data are explained by
a consistent model for grain heating and cooling.  We include a
discussion of the various infrared emission bands (IEB).
Understanding the mid IR emission is of particular interest since the
ratio of mid IR over total luminosity is a diagnostic for estimating
the relative contribution of starburst versus AGN activity (Rigopoulou
et al. 1999, Clavel et al. 2000).

Another important and so far unanswered question is how much of the
galactic infrared emission is polarised.  The observed emission in
galaxies is always averaged over a large volume and, in case
polarisation is detected, should be linked to large scale alignment of
the dust.  Polarised {\it emission} from interstellar grains can only
be detected at infrared and submillimeter wavelengths and has been
observed in various galactic sources, such as the Galactic Centre
(Hildebrand 1989), molecular clouds (Novak et al.~1989, Clemens et
al.~1999), star forming regions (Gonatas et al.~1990, Leach et
al.~1991, Smith et al.~2000), protostars (Siebenmorgen \& Kr\"ugel
2000) and prestellar cores (Minchin et al.~1995, Ward-Thompson et
al.~2000), and in the nucleus of the starburst galaxy M82 (Greaves et
al.~2000).

The galaxy NGC1808 is at a distance of 11.1\,Mpc (Reif et al.~1982),
where 1$''$ corresponds to 53\,pc.  It is classified as type Sbc
(Sandage \& Tammann 1987) and seen fairly edge--on.  The moderate far
IR luminosity of a few $10^{10}$\,\Lsun\ is ascribed by Krabbe et
al.~(1994) to young stars formed in recent star formation.  For the
central 1.5\,kpc, Koribalski et al.~(1993) suggest a total gas mass
(HI plus molecular gas) of 4.2$\times 10^{9}$\,\Msun, and Dahlem et
al.~(1990) propose 2$\times 10^{9}$\,\Msun\ of molecular gas.  In the
optical, NGC1808 reveals a disk with outer spiral arms.  Dust
filaments are seen in the North--East and seem to emerge from the
nucleus (Veron--Cetty \& Veron 1985).  There is evidence of a central
bar of 500\,pc length, an inner ring--like structure (Tacconi-Garman
et al.~1996), and a number of compact radio (Saikia et al.~1990) as
well as near IR sources (Kotilainen et al.~1996).

In Sect.~2, we present the observations followed by a discussion of
the infrared polarisation (Sect.~3).  A model for {\it astronomical
PAHs} is outlined in Sect.~4.  The infrared spectral energy
distribution is first discussed assuming optically thin emission
(Sect.~5); the model is refined in Sect.~6 to include proper radiative
transfer.  Results are summarised in Sect.~7.

\section{Observations}

\begin{figure}[htb]
\psfig{figure=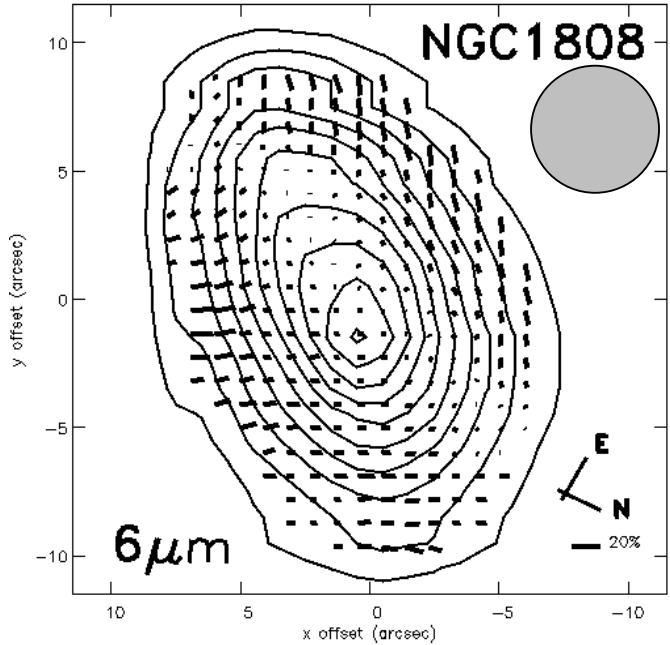,width=12.5cm}
\caption{Polarisation image of NGC1808 at 6$\mu {\rm m}$ (ISOCAM
filter lw4).  The observed E--vectors and the contours of the total
emission are shown in detector coordinates.  The celestial North-East
orientation is indicated.  Contours range from 1.28 to 8.21 in steps
of 0.77 mJy/arcsec$^2$. The effective resolution is 5$''$. }
\label{N1808_pol}
\end{figure}

ISOCAM (Cesarsky et al. 1996) polarimetric imaging on NGC1808 was set
up according to the observing template CAM05 (Siebenmorgen et
al.~1996). The corresponding ISO observation numbers, TDT, are:
87501602, 87501703, 87501804, 87501905, 87502006, 87502107 and
87502208. We used the 6$\mu$m filter lw4 and a 3$\times$3 raster with
a raster step size of 39$''$ and the 3$''$ lens.  The read out time of
each exposure was 2.1s.  After a background measurement, we took 15
exposures on the source to stabilize the detector and then a raster
through the free hole of the entrance wheel.  At each raster position
15 exposures were read out.  This procedure was carried out for each
of the three polarisers.  The polariser rasters were repeated in 2
cycles. The data were reduced with the ISOCAM interactive analysis
system (CIA version 4.0, Ott et al. 1996).  Only basic reduction
steps, such as dark current subtraction, removal of cosmic ray hits or
transient correction, are applied to the data.  They are described
together with calibration uncertainties in Siebenmorgen et al.~(2000).
The coadded images at each raster position were projected on the sky
and corrected for field distortion to derive the final mosaics.  The
mosaics have a 168$'' \times 168''$ total field of view and a pixel
scale of 1$''$.  The polarised signal is found to be consistent
between the cycles, and the average of all cycles gives the final
mosaic image for each polariser.  Our polarisation image is shown in
Fig.~\ref{N1808_pol} after correction for instrumental polarisation
(Siebenmorgen 1999).

From the ISO archive we retrieved spectro--photometric--imaging data
taken with the circular variable filter (CVF) of ISOCAM (TDT:
69702381).  For each CVF step between 16.33$\mu$m down to 5.14$\mu$m,
11 exposures of 2.1\,s integration time using the 3$''$ lens were read
out (Siebenmorgen et al.~2000, for more information on ISOCAM data
handling).  After background subtraction, we simulate multi--aperture
photometry on the images.  The spectrum, as derived for a $25''$
aperture centred on the brightest pixel, is shown in Fig.~\ref{jsm}
and Fig.~\ref{hz}.  One notices strong IEBs centered at 6.2, 7.7, 8.6,
11.3 and 12.8 $\mu$m. There is a weaker but definite broad band
emission feature centered at 7.0$\mu$m. A detection of the 7.0$\mu$m
band is reported for HD9730 by Siebenmorgen et al.~(1998) and in other
galaxies by Helou et al. (2000).  In addition there is a plateau of
fainter bands beyond 13$\mu$m and most likely we see the rise towards
the 16.4 $\mu$m band (Moutou et al. 2000), which is just outside the
scanned wavelength region. Below we combine the main bands as well as
the fainter bands and newly discovered IEBs with measured laboratory
modes of polycyclic aromatic hydrocarbons (PAH) and present
quantitative model fits.

Linear polarisation at 170\,$\mu$m was detected with the ISOPHT C200
$2{\times}2$ array (Lemke et al.~1996) by measuring several cycles
with the $0^{\circ}$, $120^{\circ}$ and $240^{\circ}$ polarisers (TDT:
85601702, 85601803, 85601904 and 85602005). The cycles were followed
by an open (sky) measurement and one of the internal reference
calibrators (Klaas et al.~1999).  We commanded 4 pointings such that
each of the 4 detector pixels measured the polarisation of the centre
position of the galaxy in succession.  In addition, the observations
on the source were bracketed by two background observations to
determine the zero signal level and to monitor instrumental
variations.  The sky measurements were processed and calibrated using
the standard procedures for photometry (Laureijs et al.~2000) with the
ISOPHT interactive analysis software (PIA version 8.1).  After
coaddition of the 4 pointings of the C200 array, we obtained a small
map which indicated that the brightness distribution of the galaxy is
not significantly different from that of a point source.  We conclude
that NGC1808 is not resolved with ISOPHT at 170\,$\mu$m.  The galaxy
is much brighter than the background and has a flux $F_{170\mu {\rm
m}} = 104{\pm}10$\,Jy, most of which comes from the central pixel.
The estimated error of 10\% is based on the ISOPHT absolute
photometric calibration uncertainty.

Since the galaxy is a point source at 170$\mu$m, we determined the
polarisation only for the four pixel--position combinations centred on
the galaxy.  The polarisation measurements of the other positions in
the map were not used as they were pointing on the wings of the beam
profile of the galaxy.  Due to the uncertainties in the beam profiles
of each detector pixel, we are not sure whether our values for
instrumental polarisation apply to measurements for which the detector
pixel is not centred on the (polarised) point source.  The processing
of the signals were carried out according to the procedure described
in Laureijs \& Klaas (1999); we also used their instrumental
polarisation values.  The resulting fractional polarisation degree
($p$) and polarisation angle ($\theta$) with respect to the equatorial
North per C200 detector pixel are presented in Table~\ref{c200pol}.
The polarisation vector obtained from pixel 3 is significantly
different from the other pixels and has a somewhat lower accuracy.
The last line in Table~\ref{c200pol} gives the polarisation derived
from the average of the Stokes vectors of all pixels after subtraction
of the instrumental polarisation.  We did not include possible
systematic uncertainties introduced by the instrumental polarisation
of the pixels.  We estimate that they amount to less than 1\% in the
final polarisation degree.

  \begin{table}[h!tb] 
	\label{c200pol}
	\begin{center} 
  \caption {Polarisation of the
  centre position of NGC1808 at 170$\mu$m for each pixel of the
  ISOPHT C200 array. Due to the 180$^{\circ}$ ambiguity, the angles in
  the first two quadrants were chosen. }

  \begin{footnotesize}
  \begin{tabular}{|c|c|c|}
  \noalign{\smallskip}
  \hline
  pixel  & $p$    & $\theta$ \\
  number & $[\%]$ & $[\rm{degrees}]$  \\
  \hline
  1 		& 4.1 $ {\pm}$ 0.9 &  75 $ {\pm}$ 6  \\
  2 		& 5.3 $ {\pm}$ 1.2 & 116 $ {\pm}$ 6  \\
  3 		& 2.2 $ {\pm}$ 1.1 &   3 $ {\pm}$ 14  \\
  4 		& 5.0 $ {\pm}$ 1.0 & 114 $ {\pm}$ 6 \\
  \hline
  {\rm mean}  	& 2.5${\pm}$0.4 & 94${\pm}$5 \\
  \hline
  \end{tabular}       
  \end{footnotesize}
  \end{center}
 
  \end{table}

\section{First estimates}

As the source is not resolved by ISO at 170$\mu$m, its far IR size
must be smaller than the diameter of the first Airy ring (140$''$).
Scans at 100$\mu$m from the KAO (Smith \& Harvey, 1996) indicate a
maximum source size of $\Theta_{\rm source} = 36''$ compatible with
the maps at 6$\mu$m (this paper), and at 8.4\,GHz and in Br$\gamma$
(Kotilainen et al.~1996).  To get an estimate of the average optical
depth, we put $F_{170\mu {\rm m}} = \Omega \cdot B_{170\mu {\rm m}}(T)
\cdot \tau_{170\mu {\rm m}}$ and assume homogeneously distributed dust
within $\Omega = 2.39 \cdot 10^{-8}$\,steradian and at temperature of 33K,
as is typical for starburst galaxies (Siebenmorgen et al.~1999).  In
this way, one gets $\tau_{170\mu {\rm m}} \leq 6.5 \cdot 10^{-3}$ in
rough agreement with the estimate by Smith \& Harvey (1996). By
applying a standard dust model (Kr\"ugel \& Siebenmorgen 1994b), this
translates into a visual optical depth $\tau_{\rm V} \sim 2000 \cdot
\tau_{170\mu {\rm m}} \leq 13$ or $\tau_{6\mu \rm{m}} \sim \tau_{\rm
V}/60 \leq 0.2$.

The fractional polarisation $p_{170\mu \rm{m}}=2.5\pm 0.4 \%$ implies
a total polarised flux of $\sim 2.5$\,Jy.  The extrapolation of the
(non--thermal) 20\,cm radio flux of $520\pm 10$\,mJy with a
$\nu^{-0.7}$--law to 170$\mu$m gives only 4\,mJy.  Therefore the far
IR polarisation of NGC1808 cannot contain any synchrotron
contribution, but must be entirely due to thermal emission from dust
grains aligned by a magnetic field.

The polarisation angle $\theta_{170\mu \rm{m}}=94^{\circ} \pm
5^{\circ}$ indicates the direction of the E--vector, which is
perpendicular to the interstellar magnetic field.  This result differs
from the strength and field direction derived from the non--thermal
radio emission.  Averaging the central 90$''$ of the 6\,cm map by
Dahlem et al.~(1990), the angles differ by almost 70\degr.  However,
both observations may probe regions in the galaxy at different
vertical heights where the magnetic field has a different horizontal
component.  A good example for such a case is presented by M\,51
(Fig.1 of Berkhuijsen et al.~1996).

Our 6$\mu$m ISOCAM map of NGC1808 also depicts the central region.
Its size down to the lowest contours equals 23$'' \times 15''$ and the
integrated flux is $1130 \pm 60$\,mJy.  A comparison with the
Br\,$\gamma$ emission (Kotilainen et al.~1996), which follows the OB
stars, shows that we also trace the starburst at 6$\mu$m.  The
starburst occurs mainly along the major axis of Fig.~\ref {N1808_pol}.
Polarisation was also detected in the ISOCAM map. The polarisation is
strongest (up to 20\%) in the outer regions and weak in the centre of
the starburst.  Along the major axis and the central region of the
galaxy where the Br$\gamma$ emission and the near IR and radio hot spots
are concentrated, the degree of polarisation is quite small.  Because
the E--vectors have a uniform direction over distances of $10''$,
which is greater than the instrumental angular resolution, the field
itself must be ordered over such a scale, corresponding to 500pc or
more.

Extrapolating the 20\,cm total radio flux towards the mid IR, we can
exclude any net polarisation from non--thermal emission.  When we
integrate at 6$\mu$m over the whole nucleus, we do not detect any net
polarisation, but obtain only a 3$\sigma$ upper limit of $5\%$.

\begin{table*}[h!tb] \begin{center} \caption {Astronomical PAH}
\label{pah.tab}
  \begin{footnotesize}
 \begin{tabular}{|c|c|c|l|l|}
\hline
  	&  	& 	&  			& \\
(1)  &  (2)	& (3)	& (4)			& (5)  \\
  	&  	& 	&  			&   \\
Wavelength & Damping Constant & Integrated Cross Section  & Mode &  Remarks \\
  	&  	& 	&  			& 	\\
$\lambda_0$ & $\gamma$ & $\sigma_{\rm int}$ &  &   \\
$\mu$m 	& $10^{12}$s$^{-1}$ &  $10^{-22}$cm$^2$$\mu$m  &  & \\
\hline
  	&  	& 	&  			&  \\
3.3  & 20	& 12	& C-H stretch		& no data \\
5.2  & 12	& 1.1 	& C-C vibration		& tentative \\
6.2  & 14 	& 21	& C-C vibration		& \\
7.0  & 5.9	& 12.5	& C-H? 			&   \\
7.7  & 22 	& 55	& C-C vibration		& \\	
8.6  & 6 	& 35	& C-H in-plane bend		& \\	
11.3 & 4  	& 36	& C-H solo out-of-plane bend & \\	
11.9 & 7	& 12	& C-H duo out-of-plane bend  &  \\	
12.8 & 3.5	& 28    & C-H trio out-of-plane bend & \\	
13.6 & 4	& 3.7	& C-H quatro out-of-plane bend & tentative  detection \\	
14.3 & 5	& 0.9	& C-C skeleton	vibration & first extragalactic detection \\
15.1 & 3	& 0.3	& C-C skeleton	vibration & tentative  detection \\
15.7 & 2	& 0.3	& C-C skeleton 	vibration & first extragalactic detection \\
16.4 & 3	& 0.5	& C-C skeleton 	vibration & data do not cover full band \\
18.2 & 3	& 1.0	& C-C skeleton 	vibration & no data \\ 
21.1 & 3	& 2.0	& C-C skeleton 	vibration & no data \\
23.1 & 3	& 2.0	& C-C skeleton 	vibration & no data \\
  	&  	& 	&  			& \\
\hline
\end{tabular}
  \end{footnotesize}
  \end{center}

(1) Center wavelength $\lambda_0$.

(3) Cross sections intregrated over the band. They are determined from
our radiative transfer model and given per H atom for C-H vibrational
modes, and per C atom for C-C modes.

(4) In the assignment of the weak bands, we follow Moutou et
al.~(1996) and Pauzat et al.~(1997).  \end{table*}

In the optical (V band), the degree of polarisation amounts to a few
percent (Scarrot et al.~1993), which is much smaller than in the mid
IR.  The polarisation vectors at V are quite regular and lie
perpendicular to the radial vectors towards the center of the galaxy.
In the inner part (750pc) of the galaxy, the polarisation is
reminiscent of a reflection pattern powered by the nucleus.

The mid IR polarisation, on the other hand, is also fairly regular,
but neither circular nor centre--symmetric.  This argues against
scattered light from the galactic nucleus.  Moreover, the grains are
tiny compared to the wavelength so that their scattering cross section
must be very small.  Therefore, the mid IR polarisation mechanism is
different from the optical and is either caused by extinction or
emission of aligned dust.  In any case, some of the interstellar
grains must be elongated and aligned, a requirement not necessary for
polarisation by scattering.

If the 6$\mu$m polarisation at a 10\% level or more is due to
extinction, the associated optical depth at 6$\mu$m must be greater
than unity (Siebenmorgen \& Kr\"ugel 2000, Fig.~11) and four or more
at 10$\mu$m.  This would imply a very strong silicate absorption
feature which is not seen.  We therefore favor the idea that the
polarisation is due to grain {\it emission}, as in the far IR.
Note that the $B$--vectors of the aligning field for the two scenarios
are perpendicular to each other.

\noindent
The polarised mid IR emission of the dust must have one of the following
origins:

\noindent
{\it 1.} It may come from PAHs heated by the interstellar radiation
field (ISRF) in the galactic nucleus (which is certainly much stronger
than the ISRF near the sun).  Indeed, our ISOCAM mid IR spectrum
(Figs.~4 and 5) shows strong IEBs. The orientation of a PAH along an
interstellar magnetic field works only if it has a magnetic moment and
if the timescale for alignment is smaller than for disorder.  Although
carbon and hydrogen atoms are not paramagnetic, Rouan et al.~(1992)
propose a number of alternative ways how to endow PAHs with a magnetic
moment.  However, their conclusion is that PAHs, whether ionized or
not, cannot align.  Therefore we do not favour PAHs to produce the
observed polarisation.

\noindent
{\it 2.} Small silicates are paramagnetic and consequently could
align. But to account for the polarisation they must also contribute
sufficiently to the emitted flux.  Calculation of the emission from
various types of transiently heated grains in reflection nebulae
(Siebenmorgen et al.~1992, 1998; Li \& Draine 2001b) show that small
silicates are not efficient radiators at 6$\mu$m.

\noindent
{\it 3.} Large grains with sizes $a \geq 100$\,\AA, must have
temperatures above $\sim$200\,K to emit substantially at 6$\mu$m.  The
heating by the ISRF is insufficient and the only possibility for them
to get so hot is close to OB stars (in so called {\it hot spots}, see
Fig.~\ref{hz_cmp} and Sect.~\ref{hotspot.sec}).  Large grains are our
favored alternative.

\begin{figure*}
\centerline{\psfig{figure=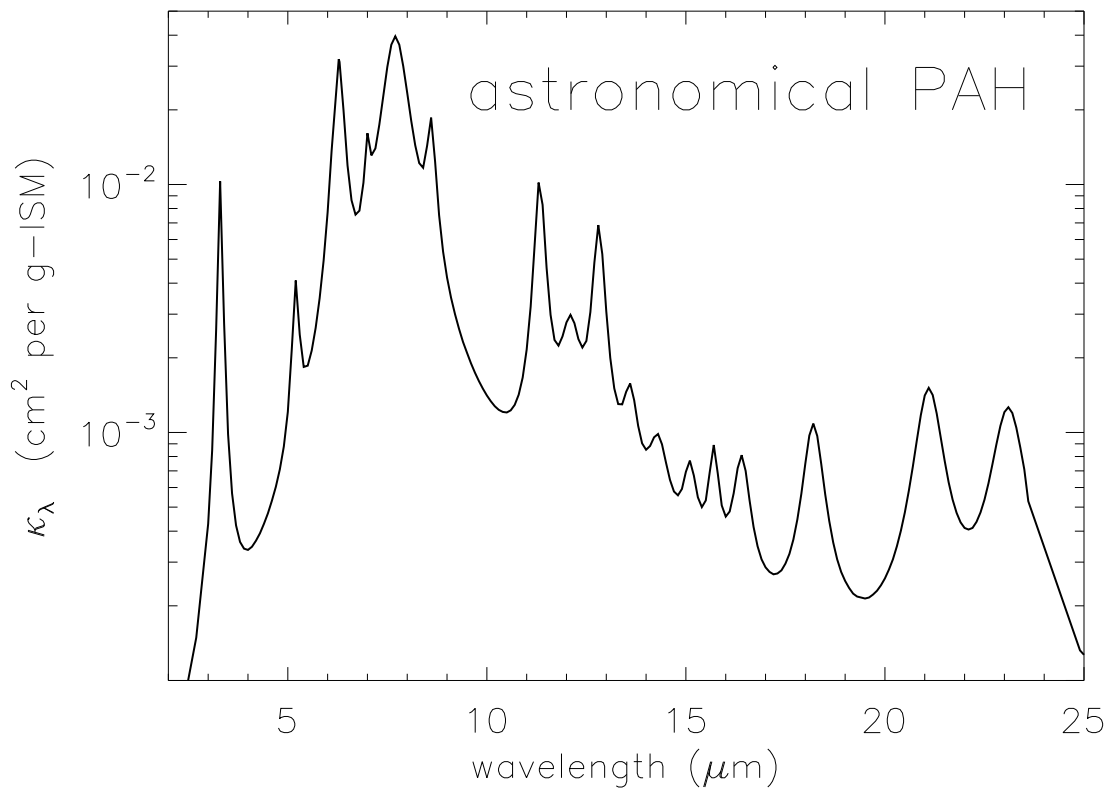,angle=90,width=16.cm}}

\caption{ Infrared absorption coefficient $\kappa_{\lambda}$ of
astronomical PAH.  There are two populations, one with $N_{\rm C}$ =
30 C atoms, the other with $N_{\rm C}$ = 500.  Both populations have
an hydrogenation parameter $f_{\rm {H/C}}$=0.2 and an abundance such
that for 10$^5$ protons in the gas phase one finds one carbon atom in
each PAH population (see also Table~\ref{pah.tab}).}

\label{kappa_IR}
\end{figure*}

\begin{figure*}
\centerline{\psfig{figure=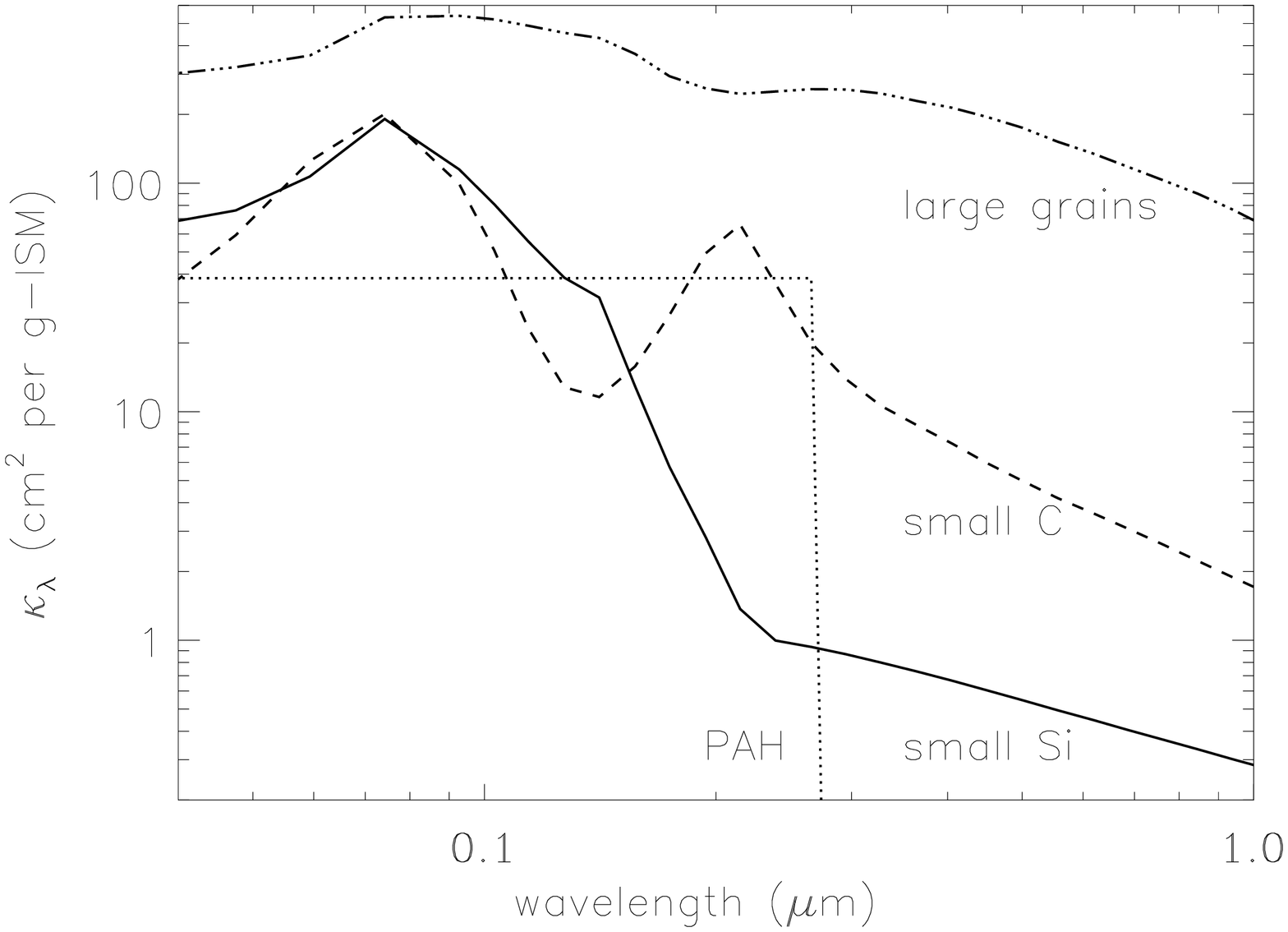,angle=90,width=17.5cm}}
\caption{Absorption coefficient $\kappa_{\lambda}$ from the UV to the near IR
for the dust model of Section 4.2: Sum of large carbon and large silicate grains
(dash--dot), small graphites (dashed) and small silicates (full line).  PAHs of
30 C atoms (dotted) with a cutoff at $\sim 0.26 \mu$m.  The absorption
coefficient of the PAHs with $N_{\rm C} = 500$ is the same, except that their
cutoff is shifted to 1$\mu$m. }
\label{kappa_UV}
\end{figure*}

\section{The dust model}

\subsection{Large and small grains}

The dust model adopted in the following two sections consists of
amorphous carbon particles (abundance in solid [C]/[H] = $2.9\cdot
10^{-4}$) and silicates ([Si]/[H] = $3\cdot 10^{-5}$) with bulk
density 2.5\,g/cm$^3$.  Their radii range from 100 to 2400\,\AA\ with
an $a^{-3.5}$ size distribution (Mathis et al.~1977).  Absorption and
scattering cross sections are calculated from Mie theory employing
optical constants from Zubko et al.~(1996) for amorphous carbon and
Laor \& Draine (1993) for silicates.

There is another dust population of very small graphites and very
small silicates with radii 5\,\AA $\leq a \leq $80\,\AA\ and size
distribution $n(a) \propto a^{-4}$ whose temperatures fluctuate.
Cross sections are again from Mie theory, optical constants for
graphites come from Draine \& Lee (1984).  The abundances of the small
graphite and small silicate grains are typically 5--10\% that of the
large carbon and large silicate grains.  The assumed abundances imply
a gas--to--dust mass ratio of $\sim$140.

\subsection{Astronomical PAH}

For simplicity, we consider only two kinds of PAHs: small ones with
$N_{\rm C} = 30$ carbon atoms and big ones with $N_{\rm C} = 500$.
Both have a hydrogenation parameter, which is the ratio of
hydrogenation to carbon atoms, of $f_{\rm {H/C}} = 0.2$ and a carbon
abundance of $Y^{\rm PAH}_{\rm C} = 3\% $ relative to the big grains.
We calculate absorption cross sections in the UV and visible following
Schutte et al.~(1993) and disregard the differences between neutral
ond ionized PAH species (De Frees et al. 1993, Onaka et al. 1999,
2000, Chan et al. 2000, Uchida et al. 2000).  For the infrared
emission bands of the C--H and C--C vibrational modes we assume
Lorentzian profiles:

\begin{equation}
\sigma_\nu \ = \ N \cdot \sigma_{\rm int} \; {\nu_0^2\over c} \cdot
 {\gamma \nu^2 \over \pi^2 (\nu^2-\nu_0^2)^2 + (\gamma \nu / 2)^2 } \ ,
\end{equation}

where $\gamma$ is the damping constant, $\sigma_{\rm int}$ the
integrated cross section and $N$ the number of carbon or hydrogen
atoms of the PAH in the particular band at center frequency $\nu_0$.
Model parameters of the PAH resonances are listed in
Table~\ref{pah.tab}.  Altogether, we consider 17 PAH bands most of
which are actually detected in NGC1808, except the one at 3.3$\mu$m
and those whose center wavelength is greater than 16.3$\mu$m.

The assignment of the feature at 7$\mu$m is unclear, it could either
be a C--C stretch or due to C--H (Allamandola et al.~1989).  There may
also be some blending with gas emission lines (e.g. [Ar~II] line at
6.983\,$\mu$m) which cannot be resolved with the CVF.  The bands with
$13.6\,\mu$m $ \leq \lambda_0 \leq 16.3\,\mu$m are seen for the first
time in an extragalactic object, but are known from Milky Way sources
(Van Kerckhoven et al.~2000, Moutou et al.~2000, Hony et al.~2001).

In the far IR, properties of PAHs are not well constrained and we
assume the continuum to fall off proportional to $\lambda^{-1.2}$ for
$\lambda \geq 24\mu$m.  Our infrared absorption cross sections are
displayed in Fig.~\ref{kappa_IR}.  They lie within the range of lab
measurements (Moutou et al.~1996, Pauzat et al.~1997, Schutte et
al.~1993) and they may be compared to other models, for example, by Li
\& Draine (2001).  Fig.~\ref{kappa_UV} shows the contribution of the
dust populations to the total absorption cross.

\begin{figure*}
\centerline{\psfig{figure=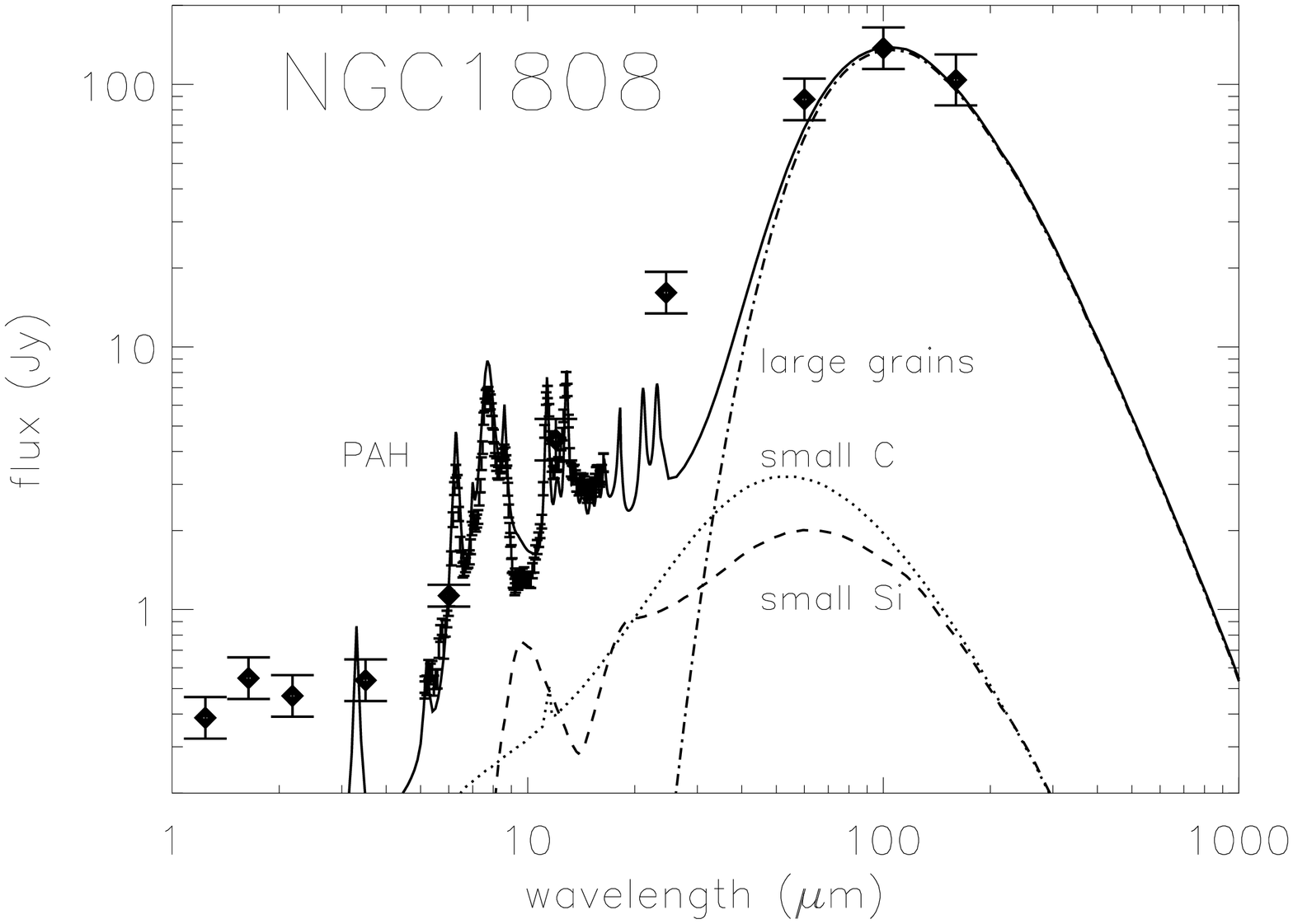,angle=90,width=16.5cm}}
\vskip -1.2cm
\centerline{\psfig{figure=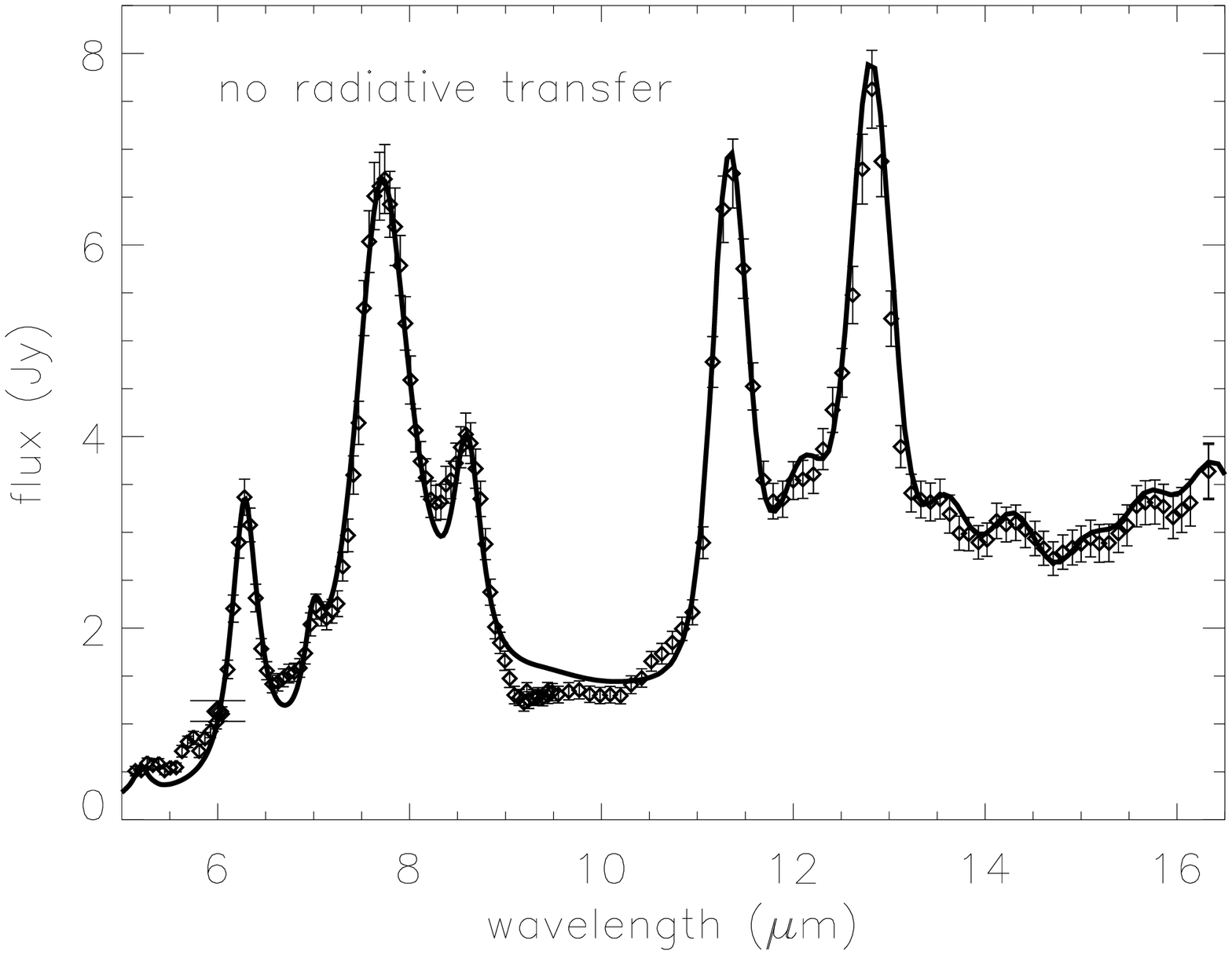,angle=90,width=16.5cm}}
\caption{{\bf Top}: Model spectrum of NGC1808 for optically thin
emission.  The large grains emit only in the far IR (dash--dot line),
small graphites (dotted) and small silicates (dashed) do not
contribute significantly to the total emission (full line).  Data
points from IRAS, ISOCAM, ISOPHT and Glass (1976).  In the mid IR, the
emission is dominated by PAHs.  {\bf Bottom:} Comparison of optically
thin model after convolution with the ISOCAM CVF spectral response
function (full line) with the ISOCAM spectrum.  }
\label{jsm}
\end{figure*}

\section{An optically thin model}

As a first attempt, we model the spectral energy distribution of
NGC1808 under the assumption that the emission is optically thin and
the dust mixture is exposed to diluted black--body radiation of mean
intensity

\begin{equation}
J_\nu = a \cdot B_\nu(T)
\end{equation}

where the dilution factor, $a$, is a free parameter.  The temperature
$T$ reflects the effective temperature of the stellar population in
the nucleus; we use $T=20000$\,K.  In the model depicted in
Fig.~\ref{jsm}, $a = 2.2 \cdot 10^{-15}$.  In this model, we have
slightly modified a few integrated cross sections of the PAHs with
respect to the numbers of column (3) in Table~\ref {pah.tab}, but
never more than 30\%.  The total gas mass, employing the abundances of
the dust components as described above, equals $M_{\rm {gas}} = 4
\cdot 10^9 \rm{M}_{\rm \odot}$.

Overall the fit to the data in Fig.~\ref{jsm} is good and for the PAH
bands even excellent.  A deficit of the model around 5.7$\mu$m could
possibly be due to yet unresolved band emission.  Such a band is
apparent in other spectra, for example, of the Orion bar and
BD+30~3639 (Fig.10 by Schutte et al.~1993).  The 5.2$\mu$m is not well
constrained by the data.  The 7.0$\mu$m IEB is detected. We point out
the blending of the 12.0 $\mu$m PAH band with the adjacent main
features.  In the optically thin model, the 6$\mu$m radiation, which
is further investigated in our polarisation study, is due to PAHs.

\section{Radiative transfer: Hot spots \label{hotspot.sec}}

\begin{figure*}
\centerline{\psfig{figure=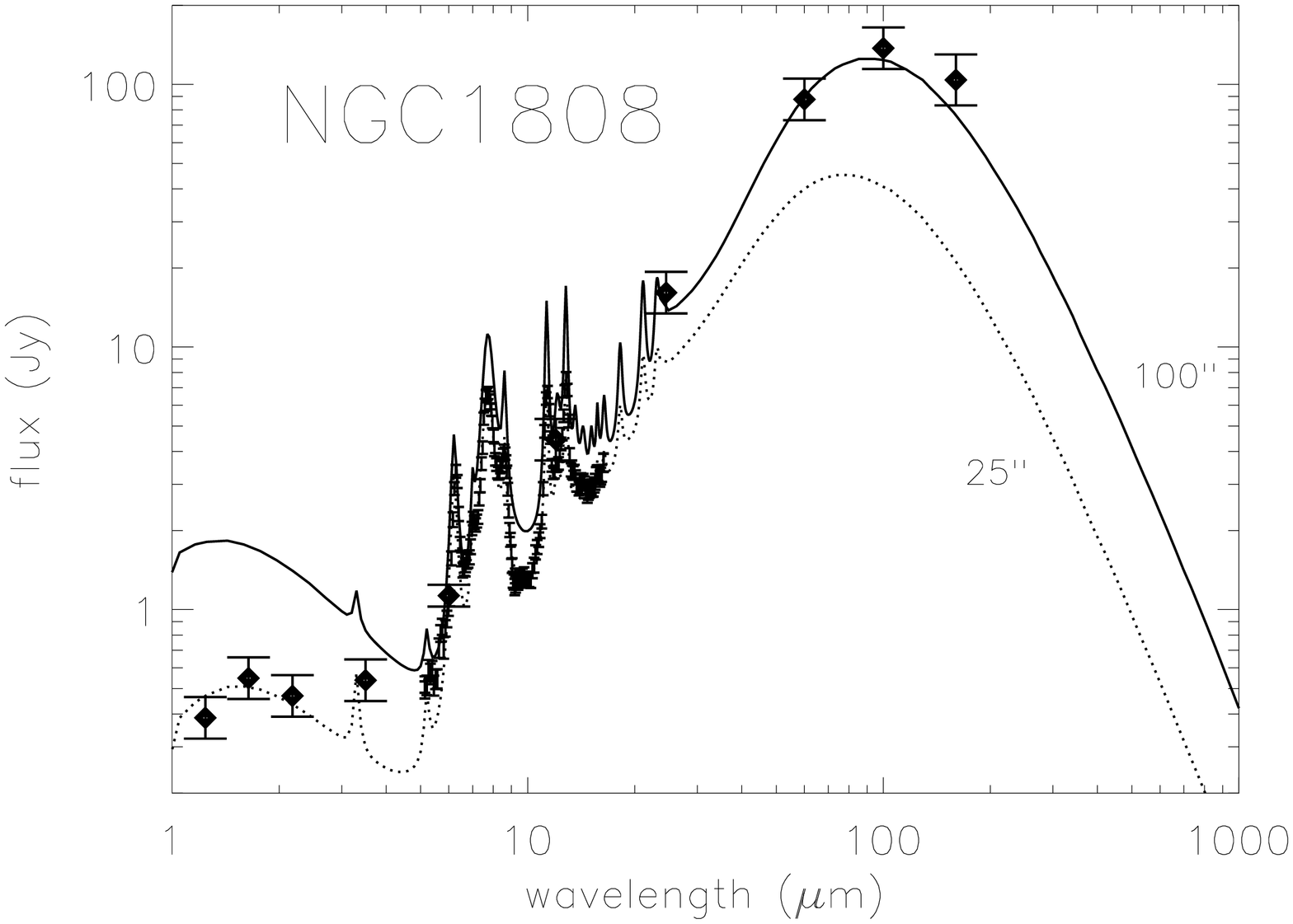,angle=90,width=16.5cm}}
\vskip -1.2cm
\centerline{\psfig{figure=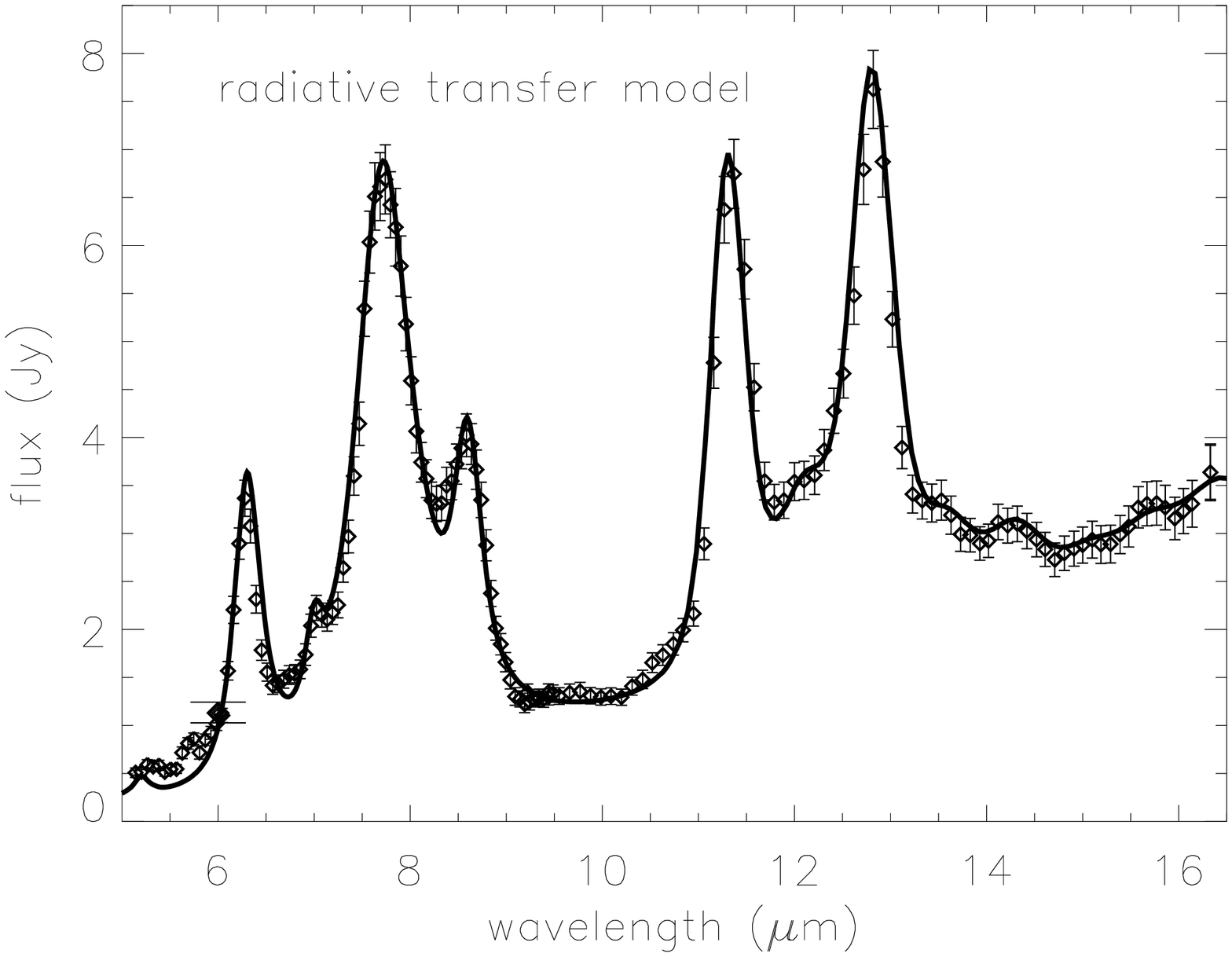,angle=90,width=16.5cm}}
\caption{{\bf Top:} Radiative transfer model spectrum of the nucleus
of NGC1808.  We show the model flux as received in a 25$''$ (dots) and
a 100$''$ aperture (solid).  Data points are shown for ISOPHT (100$''$
resolution), IRAS, ISOCAM (25$''$ aperture, barely visible), and in
the near IR (25$''$ aperture) by Glass (1976).  For model description
see text.  {\bf Bottom:} Blowup of the near and mid IR data and the
25$''$ aperture model.}
\label{hz}
\end{figure*}  

In the previous section we have matched the infrared spectral energy
distribution of the nucleus of NGC1808 by optically thin dust emission
and reasonable assumptions about the exciting radiation spectrum.
Such a model gives a good match to the CVF and far IR data,
however,

\begin{enumerate}
\item it cannot account for the 25$\mu$m emission;

\item it gives no information of the structure within the nucleus nor on the
observed IR sizes;

\item it incorrectly assumes zero optical depth within the nucleus;

\item its 6$\mu$m emission is due to PAHs.  As PAHs are unlikely to be good
polarisers, this is in contradiction to our IR polarimetric results.
\end{enumerate}

\begin{table}
\caption{Parameters of the radiative transfer model. }
\begin{center}
\begin{tabular}{|l|r|}
\hline
 &   \\
{\bf Parameter} 		& {\bf NGC~1808}\\
 &  \\
\hline
 &  \\
{\bf Distance}  		& 11.1 Mpc \\
 &   \\
{\bf  Extinction}  		&  \\ 
$A_{\rm V}$ to center ($r$=0) 	& 4.3 mag \\
 &  \\
{\bf Luminosity} 		&   \\
\ \ of all giants    L$_{\rm G}$  	& $5.5 \cdot 10^{10}$ L$_\odot$ \\
\ \ of all OB--stars L$_{\rm OB}$ 	& $0.5 \cdot 10^{10}$ L$_\odot$ \\
\ \ of one OB star   L$_{\rm HS}$  	& $10^{ \/ 4}$ \/ L$_\odot$ \\
 &   \\
{\bf Temperature} 		&   \\
\ \ of giants   		&    5000 K  \\
\ \ of OB stars 		&   30000 K \\
 &   \\
{\bf Outer radius} 			&   \\
\ \  of nucleus R$_{\rm N}$  	& 3000 pc  \\
\ \  of giants   	 	& 3000 pc \\
\ \  of OB stars R$_{\rm OB}$ 	&  350 pc \\
 &   \\
{\bf Density} &  \\
\ \  of nucleus     	& $r \leq R_{\rm OB}$: \/  \/ \quad  const \/ \\
   			& $R_{\rm OB} \leq r \leq R_{\rm N}$\/ \/ :  \quad $\propto r^{-1}$ \/ \/  \\
\ \  of giants   	& $\propto r^{-1.8}$ \\
\ \  of OB stars  	& $\propto r^{-1}$ \/ \/ \\
\ \  in hot spots 	& $10^4$ H atoms per cm$^3$ \\
 &   \\
{\bf Small PAHs} 			&  \\
\ \ abundance $Y^{\rm PAH}_{\rm C}$ 	& 6\%\\ 
\ \ size $N_{\rm C}$		& 50 \\
\ \ hydrogenation $f_{\rm{H/C}}$ &  0.2 \\
 &   \\
{\bf Large PAHs} 			&  \\
\ \ abundance $Y^{\rm PAH}_{\rm C}$ 	& 4\%\\ 
\ \ size  $N_{\rm C}$		& 500 \\
\ \ hydrogenation $f_{\rm{H/C}}$ &  0.2 \\
 &  \\
\hline
\end{tabular}
\label{para}
\end{center}
\end{table}

We therefore include the effects of radiative transfer in the nucleus
of the galaxy.  Such computations have been carried out in various
approximations by a number of authors (Rowan-Robinson \& Crawford
1989, Pier \& Krolik 1993, Granato \& Danese 1994, Efstathiou \&
Siebenmorgen 1995, Silva et al. 1998, Siebenmorgen et al.~1999b,
Efstathiou et al. 2000, Ruiz et al.~2001).  The procedure applied here
is described by Kr\"ugel \& Siebenmorgen (1994).  The model assumes
spherical symmetry which seems justified as the mid IR continuum image
of NGC1808 (Fig.~\ref{N1808_pol}) indicates an oblate structure with
ratio of major--to--minor axis of less than $3 : 2$.  It has the
following characteristics:

\begin{itemize}
\item 
There are two stellar populations both distributed within the nucleus:
{\it a)} giants of 5000K surface temperature with a radial density
distribution that falls off from the center to the edge of the nucleus
like $r^{-1.8}$.  {\it b)} OB stars ($T_{\rm OB} = 30000$K, $L_{\rm
{OB}} = 10^4$L$_\odot$ ) each of which is surrounded by a spherical
dust cloud.  Their distribution drops inversely proportional to the
radial distance from the nucleus.

\item 
There are hot spots (Kr\"ugel \& Tutokov 1978).  The heating of the
dust in the galactic nucleus does not only change on large scales, but
also locally.  A grain at a galactic distance $R$, as far as possible
away from any star, gets less heated than one at same galactic radius
$R$, but close to a star.  The hot spots are formed by the dusty
envelopes around the OB stars.
\end {itemize}

In computing the radiative transfer in the hot spots, we assume that
they are spheres.  On the surface of the spheres the dust is heated
equally by the ISRF of the galactic nucleus and by the central OB
star.  Consequently, the size of the hot spots is smaller in the
center and increases in the outer regions of the nucleus.

Note that the emission of the hot spots is not just added to the
emission from the rest of the nucleus, but is included in a
self--consistent way (see Kr\"ugel \& Siebenmorgen 1994). The main
effect of the hot spots is to drastically enhance the mid IR flux.

The model nucleus of NGC1808 has a total luminosity $L_{\rm tot} = 5.6
\cdot 10^{10}$\,L$_ \odot$, a gas mass $M_{\rm gas} = 1.2 \cdot
10^9$\,M$_\odot$, not terribly different from CO estimates (Dahlem et
al.~1990), and a radius $R\sim 3$kpc.  The dust density is constant in
the inner parts of the nucleus where the OB stars are located and
declines smoothly like $\rho(r) \propto r^{-1}$ further out.  For the
distribution of the OB stars, we take the FWHM of the 6$\mu$m image as
a guide and adopt an outer radius $R_{\rm OB} = 350$\,pc. From fitting
the near IR data, we find that the near IR flux is mostly due to giants and
put $L_{\rm G} = 5.5 \cdot 10^{10}$\,L$_\odot$.  The OB stars have a
total luminosity $L_{\rm OB} = 0.5 \cdot 10^{10}$\,L$_\odot$, each
individual OB star or hot spot has $L_{\rm HS} = 10^4$\,L$_\odot$.
The density in the hot spots is constant, $n_{\rm HS}$(H+H$_2) =
10^4$\, H atoms per cm$^3$. Model parameters are summarised in
Table~\ref{para}.

Results are shown in Fig.~\ref{hz}.  Because of the hot spots, the
25$\mu$m IRAS point is matched, which was not possible in the
optically thin model of Fig.~\ref{jsm}, and the agreement with the CVF
data is even better.

\begin{figure*}
\centerline{\psfig{figure=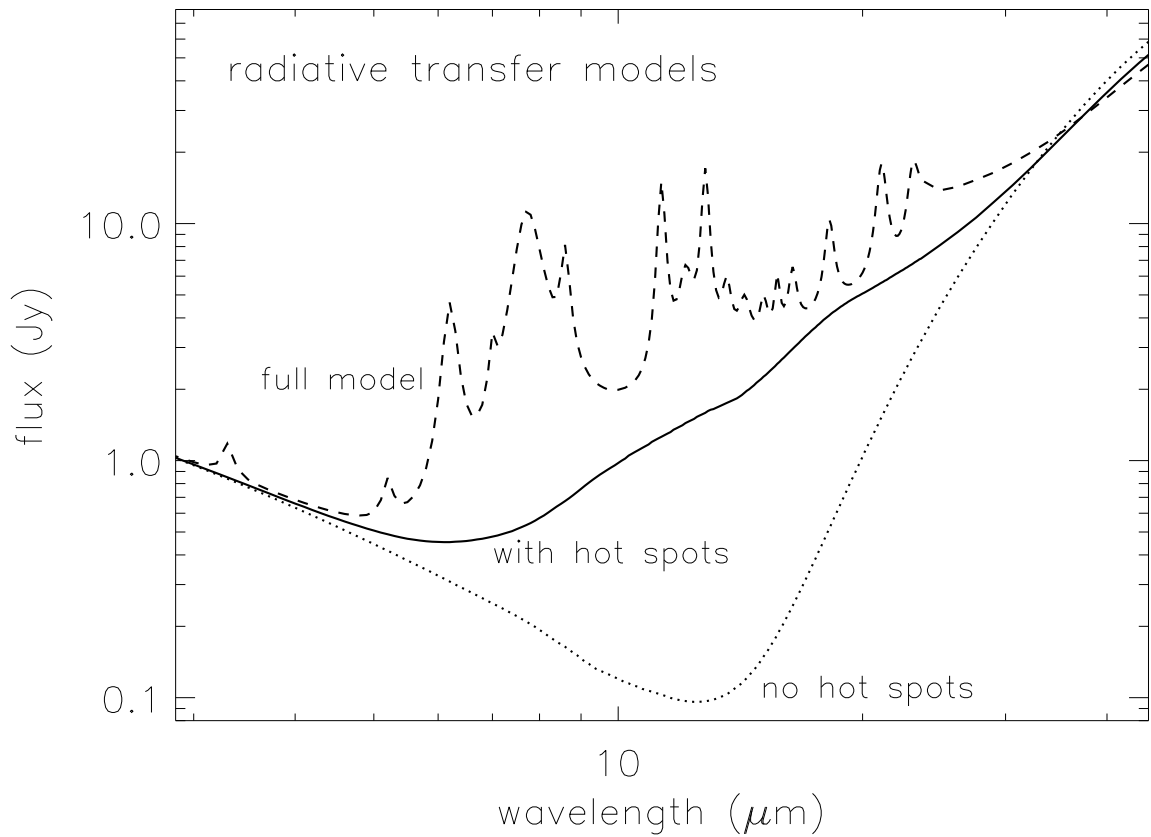,angle=90,width=17.cm}}
\caption{ To demonstrate the significance of the hot spots and of the
PAHs for the mid IR spectrum, the model of Fig.5 (dashed) is compared
to two large--grain--only models.  The first has hot spots (full line)
and indicates the continuum underlying the PAH emission; the second is
without hot spots and the dust is only heated by the ISRF.  Note the
logarithmic scale.  }
\label{hz_cmp}
\end{figure*}  

Fig.~\ref{hz_cmp} convincingly illustrates the importance of the hot
spots and of PAHs for the mid IR spectrum of a galactic nucleus.  We
mention that the contribution of the very small grains is almost
negligible; including them hardly changes anything.  Fig.~\ref{hz_cmp}
also suggests that at 6$\mu$m, the wavelength where we obtained a map
and performed the polarisation measurements, about 50\% of the flux
comes from large grains.  Therefore the proposition made at the end of
Sect.~3 that the polarisation is due to emission by large grains is
likely.

\section{Conclusion}

ISO observations of the inner $25''\times 25''$ nucleus of NGC1808
show at 6\,$\mu$m up to 20\% and at 170\,$\mu$m a net polarisation of
2.5\%. The mid IR polarisation map is the first of its kind of an
extragalactic source. To explain the mid and far IR polarisation we
rule out synchrotron radiation, scattering from large grains, PAH or
small grain emission. Emission by large ($\geq 100$\,\AA),
non--spherical grains, aligned on large scales ($\sim 500$\,pc) by
uniform magnetic fields is proposed as the major mid and far IR
polarisation driver.

ISOCAM mid IR spectroscopy revealed a multitude of emission bands. We
report the discovery of PAH features situated longward of 13$\mu$m
which are detected for the first time in a galaxy.  Based on this
observation, we derive integrated absorption cross sections of {\it
astronomical PAH} which can be used for future dust studies.

The polarimetric and spectro--photometric data are consistently
explained by a radiative transfer model which considers small
localized regions of warm dust in the immediate vicinity of early type
stars ({\it hot spots}).  We demonstrate that this local heating gives
rise to a significant contribution to the mid IR continuum by large
grains.  We find that the emission of small graphites or small
silicates is negligible.

In addition, we conclude that NGC\,1808 is not a starburst since in
our model only 10\% of the total luminosity comes from OB stars.  This
is supported by the observation that the optical depth in NGC\,1808 is
more than a factor 5 lower than in genuine starbursts like M82 or
NGC~253.

\begin{acknowledgements}
We thank the ISO operations team making it possible to complete the
ISOCAM observing sequence CAM05 on NGC1808 just a few moments before
the final shut down of the camera.  CIA is a joint development by the
ESA Astrophysics Division and the ISOCAM Consortium. The ISOCAM
Consortium is led by the ISOCAM PI, C. Cesarsky. PIA is a joint
development by the ESA Astrophysics Division and the ISOPHT
Consortium.  Contributing ISOPHT Consortium institutes are DIAS, RAL,
AIP, MPIK, and MPIA.
\end{acknowledgements}

\end{document}